\documentstyle[editedvolume,epsfig]{crckapb} 

\makeindex
\index{A:Barnes, J.E.}

\begin{opening}

\title{DYNAMICS OF MERGERS \& REMNANTS}

\author{JOSHUA E.~BARNES}

\institute{Institute for Astronomy, University of Hawai`i\\
	   2680 Woodlawn Drive, Honolulu, Hawai`i, 96822, USA}

\end{opening}

\begin{document}

\begin{abstract}
This review focus on some issues which seem relevant to recent
discussions: (1) how halo structure influences tail length, (2) the
fate of power-law density cusps, (3) the results of unequal-mass disk
galaxy mergers, and (4) the behavior of hot and cold gas in merging
disk galaxies.
\end{abstract}

\section{Halos \& Tail Length}

The \index{S:tails|(} tails of interacting disk galaxies were lucidly
explained by Toomre \& Toomre (1972).  Briefly, tails develop when
tidal forces tear galactic disks apart; material on the side of the
disk furthest from the companion galaxy, suddenly free of the
gravitational pull which had kept it in a circular orbit, escapes
along a nearly linear trajectory and so produces an ever-lengthening
tail.  Self-consistent \index{S:N-body simulations!encounters}
simulations of galaxies possessing modest dark halos have elaborated
but not fundamentally modified this basic picture (Negroponte \& White
1982, Barnes 1988).  A recent study claims that the long tidal tails
observed in many interacting systems can't escape from dark halos if
these halos \index{S:halos!dark!masses|(} have more than about ten
times the luminous mass (Dubinski, Mihos, \& Hernquist 1996).  But
this claim is model-dependent, as shown by the experiments described
below.

In these experiments, each galaxy models had three components: a
bulge, a disk, and a halo.  The bulges were Hernquist (1990) models
with mass\footnote{Unless otherwise noted, results are quoted in
arbitrary units with $G = 1$.} $M_{\rm b} = 0.0625$ and scale radius
$0.04168$; beyond a radius of $4.0$ the bulge density tapered away
smoothly.  The disks were exponentials, with mass $M_{\rm d} =
0.1875$, radial scale length $\alpha^{-1} = 1/12$, and vertical scale
height $z_0 = 0.005$.  The halos were based on the Navarro, Frenk, \&
White (1996; hereafter NFW) model, which has a logarithmically
divergent mass.  To obtain finite total masses the NFW halo profiles
were tapered:
\begin{equation}
\rho(r) = \cases{\displaystyle {M_{\rm a} \over \log(2) - 0.5} \;
			       {1 \over 4 \pi r (a + r)^2}
                   &if $r < b$\, ,\cr\noalign{\vskip 3pt}
                 \rho_{\rm b} (b/r)^2 \exp[-\gamma ((r/b)^2 - 1)]
                   &otherwise\, ,\cr}
\end{equation}
where $M_{\rm a}$ is the mass within the halo scale radius $a$ and the
parameters $\rho_{\rm b}$ and $\gamma$ were chosen so that both
$\rho(r)$ and its first derivative are continuous at $r = b$.  All the
models used halo mass and length scales $M_{\rm a} = 0.2$ and $a =
0.2$; the taper radius $b$ was used to adjust the total halo mass
$M_{\rm h}$ as follows:
\begin{equation}
\matrix{b        & 0.55285& 2.8099& 34.95\cr
        M_{\rm h}& 1.25   & 2.50  & 5.00\cr}
\end{equation}
Combining these halos with the standard bulge and disk yielded
composite models with luminous-to-dark ratios of 1:5, 1:10, and 1:20,
respectively.  All these models had identical, fairly flat
\index{S:disk galaxies!rotation curves} rotation curves out to at
least six disk scale lengths.

\begin{figure}[t!]
\vspace{7.00 true cm}
\includegraphics{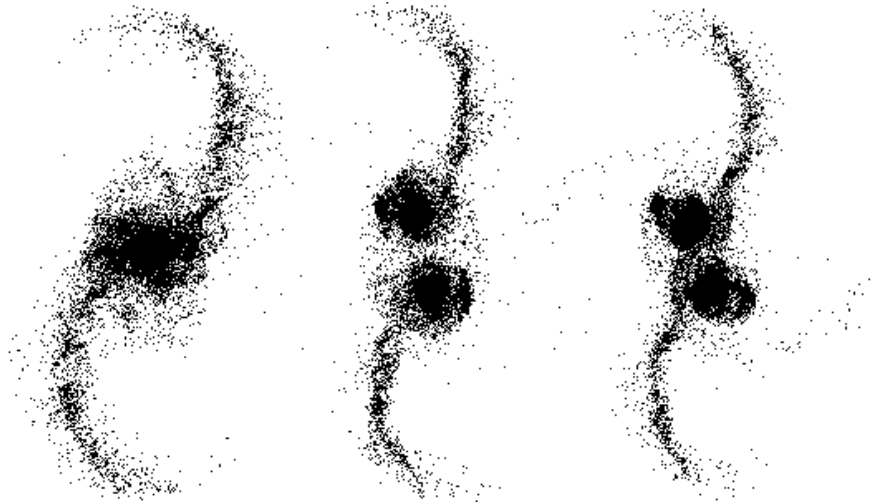}
\caption[]{Tails produced by parabolic encounters with
halo-to-luminous mass ratios of 5:1 (left), 10:1 (middle), and 20:1
(right).}
\end{figure}

All calculations started with the two galaxies only $5$ length units
apart; thus the low-mass halos were initially well-separated while the
others significantly overlapped.  The potential energy of each initial
configuration was used to assign the galaxies relative speeds
consistent with asymptotically parabolic orbits; the directions of the
initial velocities were set so as to obtain pericentric separations of
about $0.2$ length units in each case.  As shown in Figure~1, all
three experiments produced acceptable tails.

The success of this \index{S:tails|)} tail-making exercise is hardly
unexpected.  To make a proper tail, the velocity of the tail material
must exceed the local escape velocity.  The NFW model, with $\rho
\propto r^{-3}$ at large radii, has a {\it finite\/} escape velocity
even though its total mass diverges; consequently it is possible to
obtain proper tails with encounters involving arbitrarily massive
halos.  The models used by Dubinski et al., on the other hand, have
potential wells which become significantly deeper as total halo mass
increases.  {\it Long tails constrain potential well depth, but not
\index{S:halos!dark!masses|)} halo mass itself\/}.

\section{Cusp Survival}

Especially since the refurbishment of \index{S:spacecraft!HST@{\it
HST}} {\it HST\/} it's become clear that few if any early-type
galaxies have constant-density cores; instead, the \index{S:elliptical
galaxies!luminosity profiles} luminosity profiles of some
galaxies are power-laws, while others show a break and a more gradual
rise to the innermost point (eg.~Faber et al.~1997).  Core profile
shape is correlated with luminosity; bright Es have breaks, while
faint ones have power-laws.  It is natural to ask what merging will do
to this dichotomy.  A simple argument outlined below suggests that
\index{S:mergers!elliptical galaxies} merging preserves steep
power-law cusps.

Pressure-supported systems may be approximately described by
iso\-tropic distribution functions of the form $f \simeq f(E)$; in
such systems, most of the variance of the actual DF is due to its
variance with binding energy $E$.  Systems with central cusps have
singular distribution functions with formally infinite phase-space
densities; for example, a density profile $\rho \propto r^{-2}$
implies that the amount of mass with phase-space densities above $f$
scales as $M(> f) \propto f^{-1/2}$.  When systems merge,
\index{S:relaxation!violent} violent relaxation (Lynden-Bell 1967)
spreads material over a range of binding energies; subsequently,
\index{S:phase mixing} phase mixing averages phase-space density on
surfaces of constant $E$.  But this violent relaxation is {\it
incomplete\/}; \index{S:N-body simulations!mergers} numerical studies
of mergers show that potential fluctuations die down before binding
energies are completely randomized (e.g.~White 1987).  So phase mixing
can only average over a limited range of fine-grained phase-space
densities.  The final coarse-grained DF, which describes the remnant
at later times, is thus similar to the initial DF; in particular, the
DFs of remnants produced by mergers of galaxies with cusps should
still be singular after ``the dust settles''.

Merger simulations indicate that neglect of velocity anisotropy does
not badly compromise this argument.  The experiments reported here
involved parabolic collisions between identical ``gamma'' models
(Dehnen 1993; Tremaine et al.~1994), with inner density power-law
slopes of $\gamma = 1$, $1.5$, and $2$; each model had total mass $M =
1$ and half-mass radius $r_{1/2} = 0.25$.  These simulations used $N =
65536$ bodies, advanced with a leap-frog time-step of $\Delta t =
1/512$; the smoothing scale in the force calculation was $0.005$.  The
models were launched on asymptotically parabolic orbits which reached
a pericentric separation of $r_{\rm p} = 0.25$ at time $t = 1$; by
time $t = 8$ the systems had merged and largely relaxed.  Figure~2
compares the density profile of each remnant with the gamma model of
its progenitors.  As in earlier studies (eg.~White 1978; Villumsen
1983), remnant density profiles are closely related to those of the
initial systems.  Shown here more clearly than before is that steep
power-law cusps survive merging.

\begin{figure}
\vspace{5.00 true cm}
\includegraphics{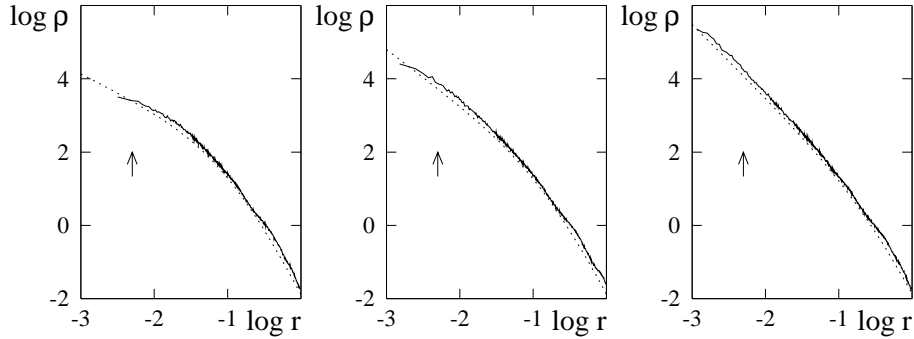}
\caption[]{Spherically-averaged density profiles of merger remnants
(solid lines) compared with initial models (dotted lines).  Results
are shown for initial models with $\gamma = 1$ (left), $1.5$ (middle),
and $2$ (right).  The arrow in each plot indicates the smoothing
length.}
\end{figure}

Such results have implications for dark halos (e.g.~NFW, Fukushige \&
Makino 1997) as well as for visible galaxies.  One implication for
galaxies is that {\it \index{S:mergers!elliptical galaxies} mergers
can't transform the power-law \index{S:elliptical galaxies!luminosity
profiles} profiles of faint Es into the broken profiles of bright Es
unless violent relaxation is somehow prolonged\/}, perhaps by the
effects of a pair of massive \index{S:black holes} black holes
(e.g.~Makino \& Ebisuzaki 1996, Quinlan \& Hernquist 1997).  If broken
profiles are produced by inspiraling black holes, the
\index{S:elliptical galaxies!centers} central regions of bright Es
should be effectively homogenized, with little velocity anisotropy and
relatively weak color and metallicity gradients.

\section{Disk Destruction}

Galactic \index{S:disks!fragility} disks are fragile; while
\index{S:mergers!minor} accretions of low-mass satellites do little
harm \index{S:mergers!disk galaxies} mergers with comparable objects
``scramble'' disk galaxies into hot spheroids.  Between these
extremes, disks may be damaged but not altogether obliterated
(eg.~Walker, Mihos, \& Hernquist 1996).
\index{S:clustering!hierarchical} Clustering models suggest that
typical mergers involve objects with broadly distributed mass ratios;
a ratio of 3:1 seems typical.  What happens to a disk galaxy which
merges with a companion one-third as massive?

A partial answer to this question emerges from a modest survey of
unequal-mass encounters (Barnes 1998).  In these experiments, both
galaxies were bulge/disk/halo systems; the larger galaxy had $3$ times
the mass of the smaller, and rotated $\sqrt[4]{3}$ times faster in
accord with the \index{S:disk galaxies!Tully-Fisher relation}
Tully-Fisher relationship.  The galaxies were launched on initially
parabolic orbits and went through several passages before merging;
remnants were evolved for several more dynamical times before being
analyzed.

Figure~3 compares the \index{S:mergers!remnants!shapes \& kinematics}
shapes and kinematics of the 3:1-mass remnants with a companion sample
of 1:1 remnants.  The plot on the left shows axial ratios, computed
from the moment of inertia of the more tightly-bound half of the
luminous material in each remnant.  As a group, the 3:1 remnants are
fairly oblate, while the 1:1 remnants are more triaxial.  The plot on
the right shows angular momentum content plotted against minor axis
ratio; the 1:1 remnants, which are flattened by velocity anisotropy,
rotate at less than half the typical rate of the 3:1 remnants.  On the
whole, the 3:1 remnants are dynamically unlike their equal-mass
counterparts.

\begin{figure}
\vspace{4.00 true cm}
\includegraphics{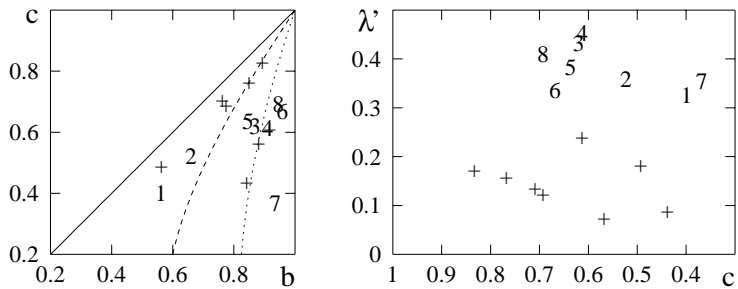}
\caption[]{Remnants of 3:1 mergers, numbered 1 to 8, compared with
remnants of 1:1 mergers, indicated by crosses.  (Left) axial ratios:
$b$ and $c$ are the intermediate and minor axis ratios, respectively.
(Right) rotation {\it vs.\/} minor axis ratio: $\lambda'$ is a
dimensionless measure of angular momentum (Barnes 1992), equal to
unity for a pure, co-rotating disk.}
\end{figure}

What \index{S:galaxy classification!Hubble sequence} Hubble type would
these objects be assigned?  Of the eight cases reported here, numbers
3, 4, 5, 6, \& 8 are fairly oblate and rapidly rotating.  Moreover,
many stellar orbits in these remnants are roughly circular; at any
given energy, most of the minor-axis tube orbits have nearly the
maximum possible angular momentum.  Finally, several of these remnants
actually look disk-like when viewed edge-on.  Apparently, {\it some
3:1-mass merger remnants are more like S0 galaxies than
ellipticals\/}.

Some \index{S:disk galaxies!early!formation} early-type disk galaxies
may thus owe their relatively hot disks and massive spheroids to
unequal-mass mergers (Schweizer, this volume).  If unequal-mass
mergers are as common as cosmological models indicate, they could form
a large fraction of S0 galaxies.

\section{Hot \& Cold Gas}

Interstellar material, though only a modest part of the total mass,
plays a profound role in \index{S:mergers!disk galaxies} disk galaxy
mergers: it reveals the large-scale kinematics of tidal features
(Hibbard et al.~1994), powers \index{S:starbursts} starbursts in
\index{S:infrared galaxies!ultraluminous} luminous IR galaxies
(Sanders \& Mirabel 1996), and builds dense \index{S:elliptical
galaxies!centers} central regions in \index{S:elliptical
galaxies!formation!mergers} early-type merger remnants (Schweizer
1990, Kormendy \& Sanders 1992).  The complex behavior of interstellar
gas is incompletely captured by the present generation of
\index{S:N-body simulations!smoothed-particle hydrodynamics} numerical
experiments.  Most simulations have in effect modeled the gas
isothermally (Barnes \& Hernquist 1996, and references therein); this
rather drastic simplification tends to obscure the relationship
between the thermal history and dynamical behavior of gas in
interacting galaxies.

\index{P:4} Color plate~4 (p.~xxii) presents a simulation which gives
some indication of the dynamics of a multi-phase medium in a merger of
equal-mass disk galaxies.  On the left are the stars, which evolved
collisionlessly.  In the middle is ``hot'' gas, assumed to be too
tenuous to cool on dynamical timescales.  On the right is ``cool'' gas
with $T \simeq 10^4 {\rm\,K}$, which serves as a proxy for molecular
and atomic interstellar material.

The bulk of the hot component roughly followed the stellar
distribution throughout the calculation.  A small amount was heated
above $10^6 {\rm\,K}$ during the first interpenetrating passage of the
two disks, but most of the gas warmed up only {\it after\/} this
passage.  When the \index{S:disks!tidal bars} tidally disturbed disks
formed bars, the resulting noncircular motions caused
\index{S:gas!shocks} shocks which heated the gas to several $10^5
{\rm\,K}$.  The final encounter of the two disks heated the gas to the
virial temperature, forming a pressure-supported atmosphere about as
extended as the stellar component (Barnes \& Hernquist 1996).  In
contrast, \index{S:spacecraft!ROSAT@{\it ROSAT}} {\it ROSAT\/}
observations of the \index{O:NGC 4038/39} Antennae (Read et al.~1995)
and other starburst systems show actual \index{S:gas!x-ray!outflow}
outflows of gas at $\sim 10^6 {\rm\,K}$; energy sources associated
with ongoing \index{S:starbursts} starburst probably power these
outflows.

Gas which cools only in the later stages of an encounter retains much
of its initial angular momentum.  Unless strong shocks develop during
\index{S:relaxation!violent} violent relaxation, this gas can't become
radically segregated from collisionless stuff; if it dissipates after
the remnant's potential settles down, it \index{S:disks!formation!gas}
form a disk rotating in the same direction as the stellar remnant.
Gradual \index{S:tails!fallback} return of gas from tidal tails
(eg.~Hibbard \& van Gorkom 1996) can build up disks with radii of many
kpc.  Since the returning gas generally falls in at an angle to the
principal planes of the remnant, such disks are likely to be warped.

Infall of \index{S:tails!fallback} material from tidal tails may have
interesting consequences even before the galaxies actually merge.  In
the right-hand column, third frame from the top, a ring of gas is seen
around the more face-on galaxy.  A video shows that this ring formed
from gas returning from the tidal tail extending to the left of this
disk (Barnes \& Hernquist 1998).  This case was particularly favorable
since the disk lay in the orbital plane, but the other disk in this
simulation, though inclined by $71^\circ$, also developed a ring.  The
extended ring of molecular gas and star formation in the northern disk
of \index{O:NGC 4038/39} Antennae \index{P:5} (Color plate~5,
p.~xxiii) may likewise have resulted from material returning from the
gas-rich southern tail; tests of this idea await more detailed
analysis and dynamical modeling of this system.

Dissipation at early stages of galactic encounters has the effect of
driving gas \index{S:gas!inflows!nuclear} inward to pool within the
central kpc of interacting disks (Icke 1985, Noguchi 1988) and at
radii an order of magnitude smaller in merger remnants (Negroponte \&
White 1982, Barnes \& Hernquist 1991).  This material must lose {\it
most\/} of its angular momentum to become so concentrated, and
\index{S:gas!inflows!gravitational torques} gravitational interaction
with collisionless material seems to be the crucial brake on the
rotation of the gas.  What little angular momentum the gas retains may
be poorly correlated with the angular momentum of the rest of the
remnant, giving rise to \index{S:elliptical galaxies!decoupled cores}
kinematically decoupled central structures (Hernquist \& Barnes 1991).

Still missing from these experiments is a proper treatment of
interactions between various phases of the interstellar material.
Simulations with \index{S:star formation} star formation and stellar
evolution are needed to incorporate \index{S:gas!feedback effects}
``feedback'' effects; one promising approach to this problem is
described by Gerritsen \& Icke (this volume).  \index{S:gas!x-ray} Hot
gas can \index{S:gas!phase transitions!ionization} ionize neutral
material as it falls back from the tidal tails; this may explain the
lack of HI in the bodies of merger remnants like \index{O:NGC 7252}
NGC~7252 (Hibbard et al.~1994).  If the pressure of the hot gas is
high enough, it may implode \index{S:gas!molecular!cloud compression}
molecular clouds, \index{S:starbursts!triggering} triggering
galaxy-wide starbursts (Jog, this volume); the extensive star
formation in the Antennae could have been triggered in this fashion.
Finally, the \index{S:gas!ram pressure} ram pressure of the hot gas
may impart significant momentum to cooler interstellar material,
thereby explaining the rather curious \index{S:tails!HI!offset}
offsets between stellar and gaseous tails observed in some interacting
systems (Schiminovich et al. 1995, Hibbard \& van Gorkom 1996, Hibbard
\& Yun 1998).

Thus, the new data from multi-wavelength studies of systems like the
Antennae offer a strong motivation for numerical simulations
incorporating both hot and cool interstellar gas.  Such simulations
are needed to interpret the observations and to test theories of
galactic transformation via violent interactions and \index{S:mergers}
mergers.  Some questions which might be answered in this way include:
What's going on in the ``overlap'' region of the \index{O:NGC 4038/39}
Antennae; is this an \index{S:interactions!collisions}
interpenetrating encounter of two gas-rich systems?  Did overpressure
of hot gas \index{S:starbursts!triggering} trigger the {\it
galaxy-wide\/} starbursts seen in this system?  Will the resulting
stars and star clusters spread throughout the
\index{S:mergers!remnants} remnant or concentrate in its central
regions?  Do systems like the Antennae give rise to \index{S:infrared
galaxies!ultraluminous} ultra-luminous IR galaxies like \index{O:Arp
220} Arp~220?  What powers the \index{S:gas!x-ray!outflow} outflows of
X-ray gas in the Antennae and in ULIR galaxies?  Can such winds clean
out the dusty central regions of ULIR galaxies, possibly exposing
central \index{S:AGN} AGNs?  What happens to the outflowing gas; how
much is returned to the \index{S:gas!intergalactic} intergalactic
medium, and how much eventually falls back?

\bigskip

\noindent
This work was supported by NASA through grant NAG 5-2836.

\begin{figure}[p!]
\begin{center}
\epsfig{figure=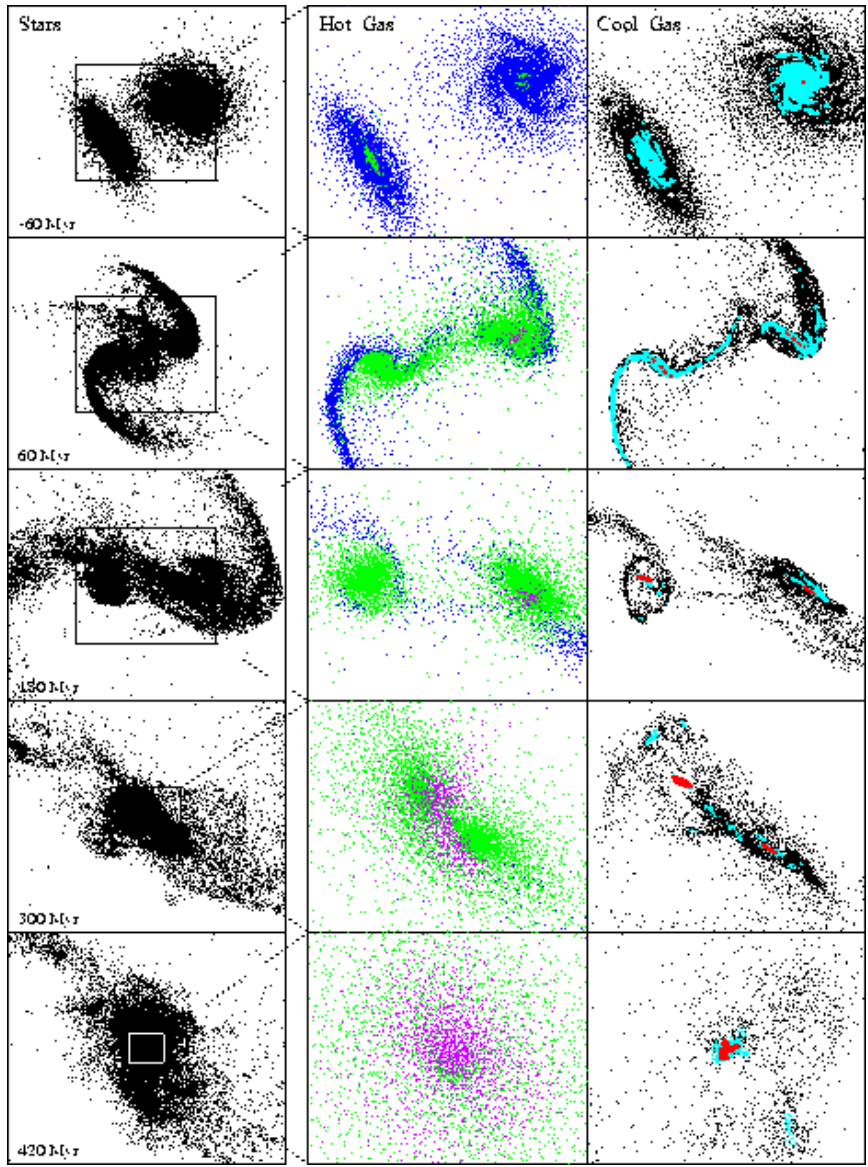,width=4.5in}
\end{center}
\caption{A parabolic encounter of two gas-rich disk galaxies.  The
stellar distribution is shown on the left; each frame is about $80
\times 96 {\rm\,kpc}$.  Times are given with respect to pericenter at
$t = 0$.  Hot gas is shown in the middle, enlarged with respect to the
stellar frame as indicated; color codes temperature, with dark blue,
green, and purple indicating factor-of-ten increases up to $2 \times
10^6 {\rm\,K}$.  Cool gas is shown on the right, on the same scale as
the hot gas; here color codes local smoothed density, with black,
light blue, and red indicating successive factor-of-hundred increases
up to $10^2 {\rm\,cm^{-3}}$.}
\end{figure}

\end{document}